\documentclass[aps,prb,preprint,
               superscriptaddress,
               preprintnumbers,
               a4paper]{revtex4}
\usepackage{amsmath}
\usepackage{amsfonts}
\usepackage{graphicx}

\begin{document}


\title{Fermi Edge Singularities in Transport through Quantum Dots}
\date{June 6, 2006}

\author{Holger Frahm}
\affiliation{Institut f\"ur Theoretische Physik, Universit\"at Hannover,
             Appelstr.\ 2, 30167 Hannover, Germany}
\author{Carsten von Zobeltitz}
\affiliation{Institut f\"ur Theoretische Physik, Universit\"at Hannover,
             Appelstr.\ 2, 30167 Hannover, Germany}
\author{Niels Maire}
\affiliation{Institut f\"ur Festk\"orperphysik, Universit\"at Hannover,
             Appelstr. 2, 30167 Hannover, Germany}

\author{Rolf J. Haug}
\affiliation{Institut f\"ur Festk\"orperphysik, Universit\"at
Hannover, Appelstr. 2, 30167 Hannover, Germany}

\begin{abstract}
We study the Fermi-edge singularity appearing in the
current-voltage characteristics for resonant tunneling through a
localized level at finite temperature.  An explicit expression for
the current at low temperature and near the threshold for the
tunneling process is presented which allows to coalesce data taken
at different temperatures to a single curve.
Based on this scaling function for the current we analyze experimental data
from a GaAs-AlAs-GaAs tunneling device with embedded InAs quantum dots
obtained at low temperatures in high magnetic fields.
\end{abstract}
\maketitle

\begin{figure}[hb]
\begin{center}
\includegraphics[width=0.25\textwidth]{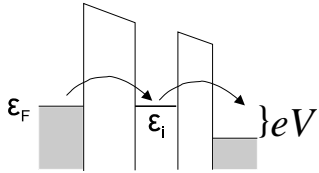}
\end{center}
\caption{
  \label{fig:schema}
  Schematic sketch of the tunneling process through a local state
  $|\epsilon_i\rangle$.}
\end{figure}

The generic response of a fermionic many body system to the sudden appearance
of a local perturbation is the Fermi-edge singularity (FES).  This response
can be observed, e.g., in experiments probing X-ray absorption in metals
\cite{Mahan67,NoDo69} or resonant tunneling through localized
levels.\cite{MaLa92} In the latter, as depicted in Fig.~\ref{fig:schema}, the
Coulomb interaction with a charge in the local state $|\epsilon_i\rangle$ acts
as a one-body scattering potential for the electrons in the leads.
The change of occupation of the local levels during the tunneling process
generates sudden changes of this scattering potential leading to
characteristic singularities in the $I$--$V$ curves at the corresponding
voltage threshold.\cite{Geim94,Nott98a,Nott98b,Dots00,GryX04,KhVd05}
A theoretical analysis of the low energy response, i.e.\ in the vicinity of
these singularities, at zero temperature is possible using bosonization
techniques.\cite{Scho69} In this approach the response functions can be
expressed as correlation functions of boundary changing operators in an
equivalent 1+1 dimensional conformal field theory.\cite{AfLu94}
For the tunneling current in the vicinity of the voltage threshold
$V_0$ this gives a power law,
\begin{equation}
   I(V,T=0) \propto (D/e(V-V_0))^\gamma\,\theta(V-V_0)\ ,
\label{IV0}
\end{equation}
with a characteristic high energy cutoff $D$ of the order of the band width.
The critical exponent $\gamma$ at this threshold is completely determined by
the scattering potential and can be expressed in terms of the scattering phase
shifts in the channels coupled to the scatterer taken at the associated Fermi
momenta.\cite{Scho69,NoDo69,MaLa92}
The appearance of power laws for the response of the system in this approach
is a consequence of the absence of an intrinsic scale in the low-energy theory
for this problem.

There are several effects, however, which affect this singularity
and therefore obstruct their direct observation:
already at zero temperature the finite lifetime of an electron in the local
state leads to an intrinsic broadening of this level.  Since its energy
$\epsilon_i$, measured relative to the Fermi energy, depends on the voltage as
$\epsilon_i\propto e(V-V_0)$ this broadening directly modifies
Eq.~(\ref{IV0}).\cite{MaLa92}
In addition, nonequilibrium effects causing dissipation have to be taken into
account for a complete description of the system with a voltage bias across
the tunneling barrier. This implies further smoothing of the FES due to the
resulting finite lifetime of states in the leads.\cite{BHGS00,MuAB03,AbLe05}
Even without dissipation the response functions such as $I(V)$ of the system
are exponentially damped at finite temperature, i.e.\ away from criticality.
However, as the temperature is lowered, Eq.~(\ref{IV0}) has to be approached,
which is reflected by a divergence of the maximum of the measured current as
$I(V_{\max},T)\propto T^{-\gamma}$, again limited by the intrinsic broadening
of the local level.  Here, the critical exponent $\gamma$ is the same as the
one governing bias dependence of the $T=0$ response.
Finally, the derivation of the expression (\ref{IV0}) for the
tunneling current in the bosonization approach relies on a linear
dispersion of the lead electrons.  While this approximation is
valid close to the threshold, nonlinearities or fluctuations of
the density of states (DOS) due to the presence of impurities in
the leads will modify the current for finite detuning $\Delta V =
V-V_0$ of the voltage.  These corrections are a property of the
probe and independent of the temperature.  Therefore, they can be
distinguished from the FES if its dependence on both temperature
\emph{and} voltage, at least at low energies, is known.

Neglecting other than thermal broadening a closed expression for this
dependence can be derived for lead electrons with linear dispersion: at
finite temperature $T$ and at a finite detuning $\Delta V$ of the voltage from
the threshold the response of the system will depend on the ratio $e\Delta
V/k_BT$ of these energy scales reproducing the power laws mentioned above for
zero temperature and at the maximum, respectively.
%
%
For non-interacting electrons the extension of Eq.~(\ref{IV0}) for the X-ray
absorption problem to finite temperatures can be achieved by taking into
account the Fermi distribution for the initial states \cite{OhTa84,OhTa90}
(see also Ref.~\onlinecite{YuAn70}).  Alternatively, it can be obtained by
incorporating temperature in the bosonization procedure of
Ref.~\onlinecite{Scho69}.  Within the latter approach the analysis is easily
generalized to the case of \emph{interacting} 1D fermions described by the
Luttinger Hamiltonian \cite{AfLu94,GoNeTs:boson}: the Luttinger parameter
$K_\rho$ only enters the exponent $\gamma$ in the final expressions
(\ref{IV0}) and (\ref{IVT}).
To be specific, we consider a single channel of non-interacting fermions in
the emitter of the tunneling device coupled to a local level at the origin.
The problem can be formulated as a one-dimensional scattering problem in terms
of a left moving (chiral) fermion field $\psi$ on the interval $[-L,L]$ with
Fermi velocity $v_F$ described by the Hamiltonian (see e.g.\
Ref.~\onlinecite{AfLu94})
\begin{equation}
  {\cal H} = \int\mathrm{d}x\,\left\{iv_F \psi^\dagger \frac{d}{dx} \psi
  + \delta(x) U \psi^\dagger\psi\,b_i^\dagger b_i\right\}
  + \epsilon_i b_i^\dagger b_i
\label{ham1}
\end{equation}
($b_i$ annihilates the electron in the local state).
Using Fermi's golden rule with the tunneling operator $\left(\psi^\dagger(0)
b_i+h.c.\right)$ the current can be expressed in terms of the Green's function
\begin{equation}
  G(t) = \left\langle \left({\cal U}\psi(0,t)\right)^\dagger
           {\cal U}\psi(0,0)\right\rangle_T\ .
\label{GF}
\end{equation}
Here, ${\cal U}(t)$ is the unitary operator describing the change of the
boundary condition at $x=0$ due to the tunneling of an electron.  It relates
the Hamiltonian (\ref{ham1}) in the sectors with the local level occupied or
empty by means of a canonical transformation.\cite{Scho69}
Representing the fermion $\psi\propto\exp(-i\sqrt{4\pi}\phi)$ in terms of a
left-moving boson $\phi(x,t)$ with mode expansion (the zero mode $\phi_0$
does not contribute to (\ref{GF}))
\[ 
  \phi(x,t) = \phi_0 + \sum_{m=1}^\infty \frac 1 {\sqrt{4\pi m}} \left(a_m
   \mathrm{e}^{-i m \pi (x + v_F t) /L}
        + h.c. \right)
\] 
\begin{widetext}
\noindent
the Green's function is ($\delta=U/2v_F$ is the Fermi phase shift)
\[ 
\begin{aligned}
  &G(t) = \exp \left\{ \left(1 + \frac{\delta}{\pi}\right)^2
  \sum_{m=1}^\infty \frac{1}{m}
        \left[ 2 \left< a_m^\dagger a_m \right>_T (\cos{(m\pi t /L)} -1) +
  \mathrm{e}^{-i m \pi t/L} - 1 \right] \right\}
\end{aligned}
\] 
\end{widetext}

For $T=0$ the ground state expectation value of the bosonic occupation numbers
vanishes, $\langle a_m^\dagger a_m\rangle_0=0$, and this expression leads
directly to (\ref{IV0}).\cite{Scho69,AfLu94} At finite temperatures the
Bose-Einstein distribution of the bosonic occupation numbers $\langle
a_m^\dagger a_m\rangle_T$ is used instead.\cite{EgMK97,MaEJ97}
Within this approach we obtain ($\gamma=-2\delta/\pi-(\delta/\pi)^2$)
\begin{equation}
\begin{aligned}
    &I(\Delta V,T) \propto\int_0^\infty \mathrm{d}t\,
        {\rm e}^{i e\Delta V t}\,G(t)
\\
    &\quad= \frac{1}{\pi}\mathrm{Im}\left[
        \left(\frac{i \beta D}{\pi}\right)^\gamma
        \mathrm{B}\left(\frac{1-\gamma}{2} -i \frac{\beta e\Delta
        V}{2\pi}, \gamma \right) \right]
\end{aligned}
\label{IVT}
\end{equation}
for the low temperature response (see also Refs.~\onlinecite{OhTa84,OhTa90}).
Here $\beta=1/k_BT$ and $\mathrm{B}$ is the beta function.

This expression reproduces the power laws mentioned above:
(i) at voltages sufficiently above the threshold, for $e\Delta V\gg k_BT$,
the power law singularity Eq.~(\ref{IV0}) emerges from Eq.~(\ref{IVT}).
(ii) For positive $\gamma$, the tunneling current becomes maximal at some
finite value of $\beta e\Delta V$ which approaches $0$ for $\gamma\to1$ and
diverges as $\gamma\to0$.  Since a given FES is described by a $T$-independent
exponent $\gamma$, this implies that the position of the maximum in the
$I$-$V$ curve moves to higher voltage with the temperature
$V_{\max}-V_0\propto T$.  At this maximum, the only temperature
dependence of the current is in the pre-factor, i.e.\
$I_{\max}\propto T^{-\gamma}$.
(iii) Finally, for $\gamma\to0$, i.e.\  no scattering of the
electrons off the local level irrespective of its occupation,
Eq.~(\ref{IVT}) reduces to a thermally broadened current step $I(\Delta V,T)
\propto 1/\left(1+\exp\left(-\beta e\Delta V\right) \right)$.
%

In the following we apply this result to the analysis of edge
singularities observed in tunneling experiments through localized
levels in InAs quantum dots in a strong magnetic field.
These self-assembled InAs quantum dots are sandwiched between two AlAs
tunneling barriers. They have a height of approximately 3\,nm and a diameter
of $2r_0\approx10-15\,\mathrm{nm}$\cite{Dots00}. The emitter and collector
consist of a 15\,nm thick undoped GaAs spacer layer followed by a $n$-doped
GaAs buffer with graded doping: a 10~nm thick layer of $\mathrm{n^{-}}$ doping
($1\cdot 10^{16} \mathrm{cm}^{-3}$), a 10~nm thick layer of n doping ($1\cdot
10^{17} \mathrm{cm}^{-3}$) and a 1~$\mu$m thick layer of $\mathrm{n^{+}}$
doping ($2\cdot 10^{18} \mathrm{cm}^{-3}$).
We have studied two samples with nearly the same growth parameters; only the
tunneling barrier thicknesses are different. One has AlAs tunneling barrier
thicknesses of 4 and 3\,nm and we show here measurements at a magnetic field
of 14.9\,T. The other sample with more asymmetric barriers (5 and 2\,nm) was
studied at magnetic fields up to 28\,T. In this paper we look at the tunneling
direction where the electrons tunnel first through the thicker barrier onto
the dot and leave it through the thinner barrier.

The thickness of the spacer layer (15\,nm) of the investigated samples was
carefully chosen to allow for a description of the emitter as a three
dimensional electron gas even at the presented bias voltages, different from
the structures used in Ref.~\onlinecite{Nott96}. A comparison of the $I$-$V$
characteristics to a test sample with a much thicker spacer layer (100\,nm)
showed distinct differences. For this test sample strong peaks in the $I$-$V$
characteristic due to the formation of 2D subbands in the emitter right before
the tunneling barriers evolved.  In Ref.~\onlinecite{Nott96} such a 2D emitter
behavior was also observed for the same 100\,nm thickness of the spacer layer.
The absence of such peaks in all our samples with a 15\,nm spacer layer
confirms our assumption of a 3D emitter.  Additional support fro this
assumption has been given by a detailed analysis of a sample of the wafer with
the 4 and 3\,nm barriers in Ref.~\onlinecite{Nauen04}.  Both the current and
the measured shot noise showed excellent agreement with the theoretical models
for 3D-0D-3D tunneling.

\begin{figure}[ht]
\begin{center}
\includegraphics[width=0.45\textwidth]{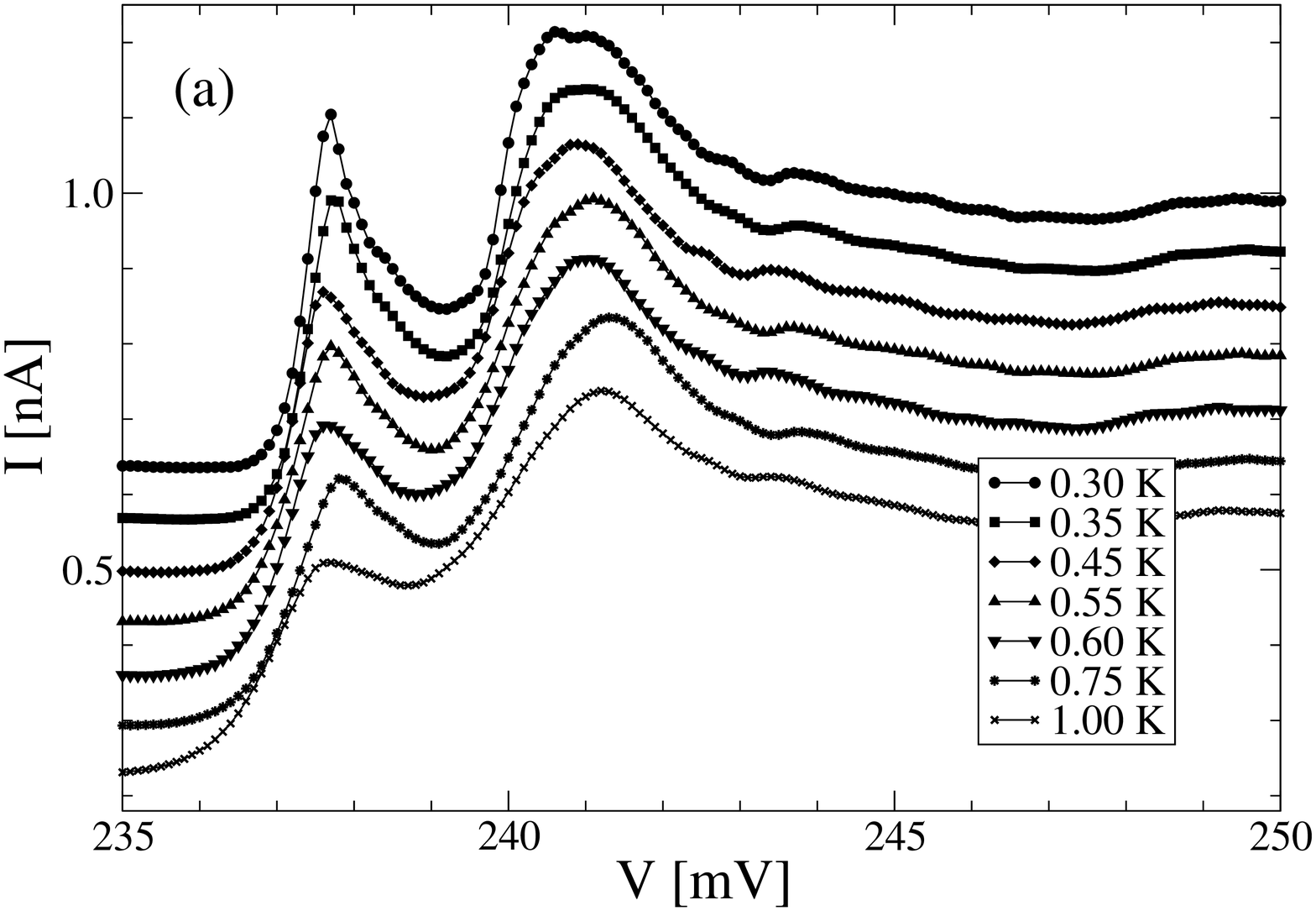}

\includegraphics[width=0.45\textwidth]{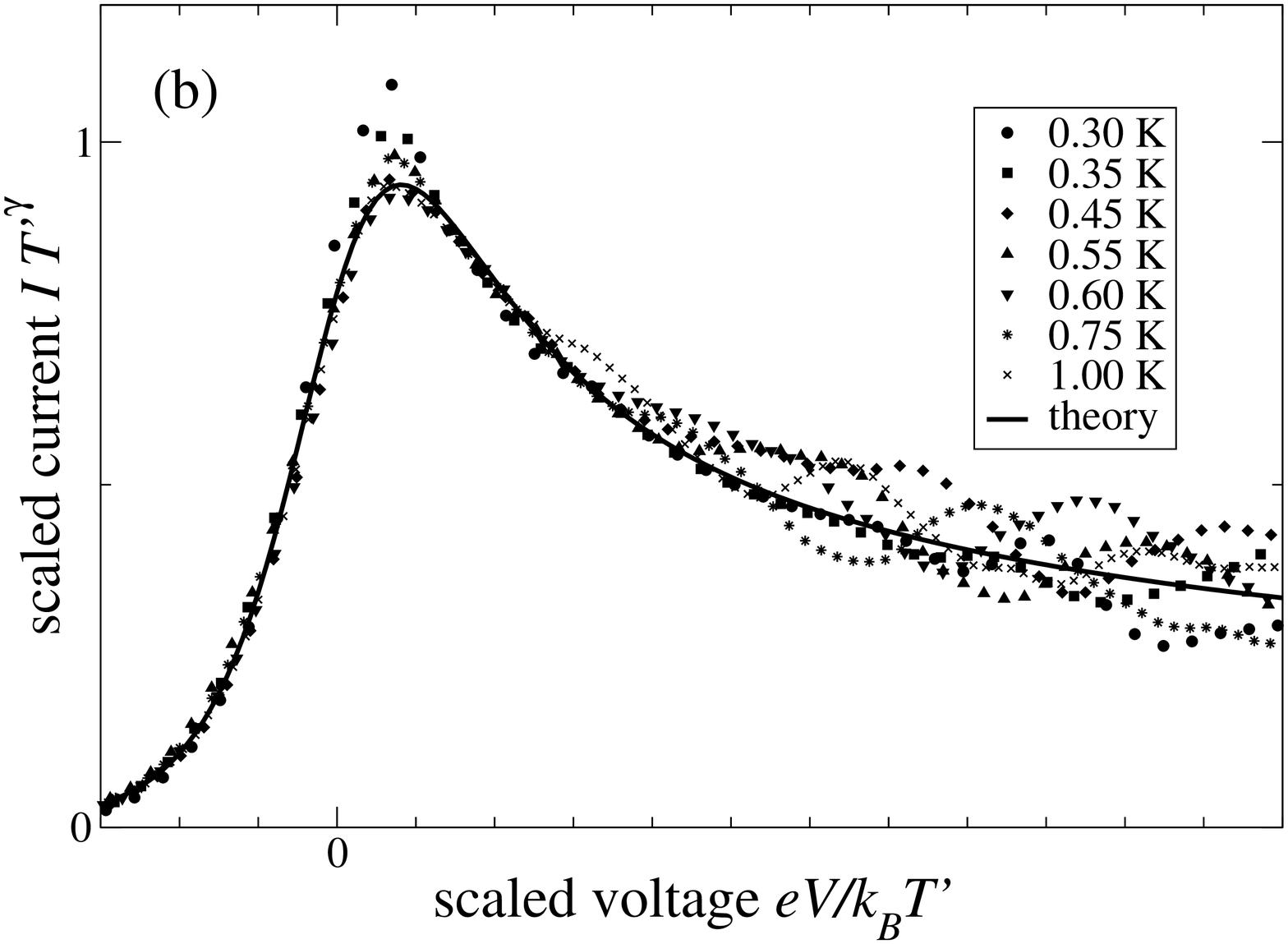}
\end{center}
\caption{
  \label{fig:edge_coll}
  (a) $I$-$V$ characteristics exhibiting the spin-split Fermi edge singularity
  for a magnetic field of $14.9\,T$ at temperatures $T=0.3\,K$, $0.35\,K$,
  $0.45\,K$, $0.55\,K$, $0.6\,K$, $0.75\,K$, $1.0\,K$ (for clarity a
  $T$-dependent offset is added to $I$).  (b) Same data for the FES at
  $V\approx238\,mV$ with current rescaled as $I\,T'^\gamma$ vs.\ voltage in
  units of the effective temperature $T'$ (see main text) for a collapse onto
  a single curve.  The full line is the best fit of Eq.~(\ref{IVT}) to the
  data.}
\end{figure}

\begin{figure}
\includegraphics[width=0.4\textwidth]{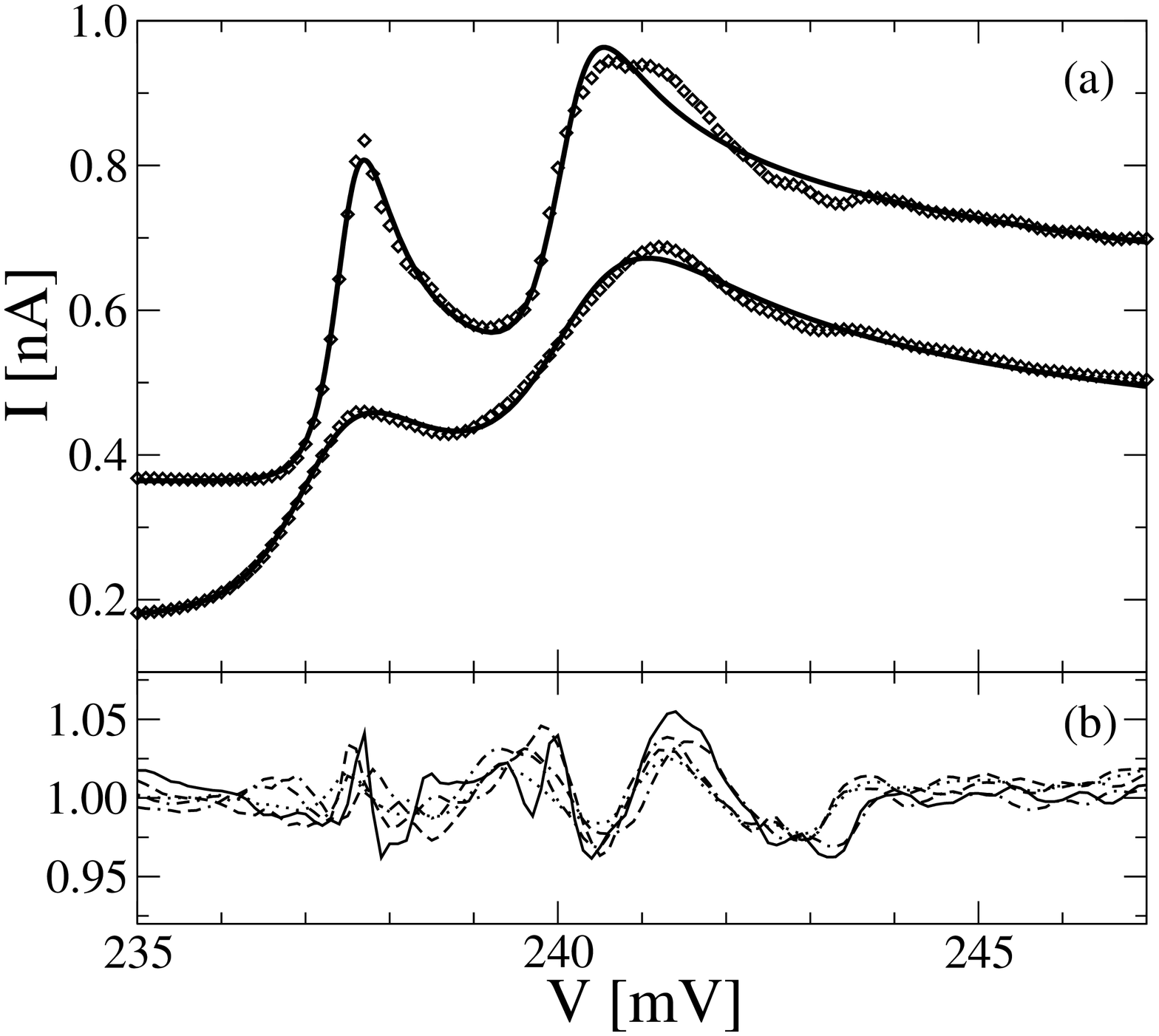}
\caption{
  \label{fig:IVCurve}
  (a) $I-V$ data for a Zeeman-split edge singulartiy at $B=14.9\,T$,
   $T=0.3\,K$ (upper curve) and $T=1.0\,K$ (lower curve). The dots show the
   experimental data, the solid line is the theoretical fit. (b) Ratio of
   experimental data and theoretical fit for temperatures $T=0.3\,K$,
   $0.45\,K$, $0.6\,K$, $0.75\,K$, $1.0\,K$ showing the (essentially
   $T$-independent) fluctuations in the local DOS of the emitter.  }
\end{figure}

Expression (\ref{IVT}) for the tunneling current implies that the $I$--$V$
data taken at different temperatures $T$ -- after rescaling the current as
$IT^\gamma$ and the voltage as $e\Delta V/k_BT$ -- should lie on a single
scaling curve.  Here, the only free parameter should be the edge exponent
$\gamma$.  In practice, however, there are various additional effects to be
considered: first, only a fraction $\alpha$ of the total applied voltage drop
occurs between the emitter and the quantum dot, i.e.\ $\epsilon_i=\alpha
e\Delta V$.  The energy-to-voltage conversion factor $\alpha$ can be
determined from the thermal broadening of the FES.  Second, as already
mentioned in the introduction, the local level in the quantum dot has an
intrinsic width $\Gamma_i$ due to its hybridization with the states in the
emitter.  
The effect of these mechanisms on the broadening of the $I$-$V$ curve is
different and non-equilibrium conditions\cite{MuAB03,AbLe05} have to be taken
into account for the latter.  Here we include both by introducing an effective
temperature $k_B T' \equiv \sqrt{(k_BT)^2+\Gamma_i^2}$ in Eq.~(\ref{IVT}).
Based on the theoretical expression we have analyzed data taken with the first
sample in a magnetic field of $14.9\,T$ at various temperatures between $0.3$
and $1.0\,K$ (Fig.~\ref{fig:edge_coll}(a)).  Due to the Zeeman-splitting of
the local level in the dot the FES appear in pairs (see also
Refs.~\onlinecite{Nott98,Nott98b}).  Hence the data have to be described by
the sum of two contributions (\ref{IVT}) with different edge exponents
$\gamma_\uparrow$, $\gamma_\downarrow$ but the same conversion factor $\alpha$
and intrinsic broadening $\Gamma_i$.  Unlike in bulk InAs the Land\'e factor
of the quantum dots is positive.\cite{Nott98,Dots00} As a consequence we
identify the FES at lower bias voltage with tunneling of spin-$\downarrow$
electrons.
Performing a fit to the experimental data we obtain $\gamma_\uparrow=0.27$,
$\gamma_\downarrow=0.46$, $\alpha=0.15$ and $\Gamma_i=0.36\,K$.
Now, the FES from tunneling of electrons with spin $\downarrow$ (corresponding
to the peak in the $I$-$V$ curves at lower bias) can be isolated by
subtracting the theoretical contribution from the other spin projection.
Rescaling the axes as described above the $I$-$V$ data are indeed seen to
collapse onto a single curve given by (\ref{IVT}) (see
Fig.~\ref{fig:edge_coll}(b)).  The scatter observed at higher voltages is
caused by fluctuations in the local DOS in the emitter.
As argued above these fluctuations should lead to
temperature independent deviations of the experimental data from the
theoretical prediction.  This behaviour is shown in Fig.~\ref{fig:IVCurve}
where the raw $I$--$V$ data (i.e.\ without rescaling of the axes) near the
Zeeman-split FES at two different temperatures are shown.  Taking the ratio of
the experimental data to the theoretical fit the temperature independence of
the fluctuations in the current becomes manifest, see
Fig.~\ref{fig:IVCurve}(b).  Although the data show some scatter which is most
likely caused by noise in the experimental data, it is astonishing that in all
five experimental curves the same small deviation in the theoretical curve can
be seen.

As has been reported before,\cite{Dots00} the edge exponents characterizing
the FES in the tunneling current depend strongly on the magnetic field.
To understand this in terms of the parameters describing the dot and its
interaction with the emitter we have to go beyond the generic low-energy
Hamiltonian (\ref{ham1}): the broadening of the localized level and the Fermi
phase shifts determining the edge exponents in (\ref{IV0}), (\ref{IVT}) need
to be computed within a suitable microscopic model.
As discussed above the emitter can be described
as a 3D electron gas in the half space $z<0$ (see, e.g., Ref
\onlinecite{WCBM98} for the description of an effective 2D electron system).
In the experiment the transverse motion of the electrons is quantized by the
magnetic field $B||\hat{z}$.  Only the lowest Landau band (LLB), $n=0$, is
occupied and the electrons populate one-dimensional channels of given angular
momentum $m\ge 0$ with Fermi momentum $\hbar k_{F\sigma}$ for spin projection
$\sigma=\uparrow,\downarrow$.\cite{Dots00}
The radial wave functions in these channels, $\psi_{n=0,m}(\rho,\phi) \propto
\rho^m\exp(-im\phi-\rho^2/4\ell_0^2)$, depend on the magnetic field through
the magnetic length $\ell_0=\sqrt{\hbar/eB}$.  In the experiments, $\ell_0$ is
comparable to the lateral size $r_0$ of the quantum dot.

Assuming a Gaussian wave function for the electron on the isolated quantum dot
the intrinsic broadening $\Gamma_i$ can be computed in perturbation theory
from its overlap with the single electron states in the LLB of the leads (in
the present geometry the broadening will be dominated by the overlap with the
collector states).  The field dependence of the wave functions $\psi_{0m}$
leads to a linear growth of the broadening with the magnetic field with a
change in slope around $\ell_0\approx r_0$.  This agrees with the qualitative
behaviour of $\Gamma_i$ as obtained from our fits to the experimental $I$-$V$
data: analyzing the evolution of the FES with the field we find that
$\Gamma_i$ varies between $0.36\,K$ at $B=14.9\,T$ and $1.5\,K$ at $B=28\,T$.
as a consequence, the corresponding lifetimes of particles in the local state
are well above the thresholds for the observation of the FES.\cite{Nott00}.

Finally, we want to discuss the magnetic field dependence of the edge
exponents $\gamma_{\sigma}$ characterizing the Zeeman-split FES.
In the tunneling experiment
the scattering potential of the dot affects \emph{several} channels labelled
by the angular momentum $m$ and spin $\sigma$ of the electrons.  Alternation
of the occupation of the quantum dot implies changes $\delta_m(k)$ in the
Fermi phase shifts experienced by the electrons in channel $m$.  As in the
simplified model (\ref{ham1}) the edge exponents can be expressed in terms of
these changes (angular momentum conservation implies that only $m=0$ electrons
contribute to the tunneling matrix element) \cite{NoDo69}
\begin{equation}
  \gamma_\sigma = -\frac{2}{\pi}\delta_0(k_{F\sigma})
  -\frac{1}{\pi^2}\sum_{m,\sigma'} \delta_{m}(k_{F\sigma'})^2\ .
\label{gamma}
\end{equation}
The scattering potential for the electrons in the emitter is composed from the
band bending due to the presence of the semiconductor-insulator interface at
$z=0$ and the screened Coulomb potential of a charge on the
quantum dot -- if it is occupied.  In the 1D Landau channels with given
angular momentum $m$ both can be described by effective potentials
$v\mathrm{e}^{\kappa z}$ decaying exponentially into the emitter with range
$\kappa^{-1}$ being the Debye radius.
The resulting phase shift picked up by the 1D electrons with momentum $\hbar
k$ perpendicular to the boundary is
\begin{equation}
  \delta(k) = -\frac{i}{k} \ln\left( \frac
        {{_0F_1}\left(;1+\frac{2i k}{\kappa},\frac{v}{\kappa^2}\right)}
        {{_0F_1}\left(;1-\frac{2i k}{\kappa},\frac{v}{\kappa^2}\right)}
  \right)
\label{deltaexp}
\end{equation}
where $_0F_1$ is a hypergeometrical function.  For the empty dot the effective
potential $v$ is the band bending $v_b$ depending on the doping profile.  For
the charged dot $v=v_b+v_{n,m}$ where $v_{n,m}$ is obtained by projecting the
screened Coulomb potential $V(\rho,z)$ (see e.g.\ Ref.~\onlinecite{MaLa92}) to
the radial wave functions $\psi_{n,m}(\rho,\phi)$ of the electrons in the
$n^{\mathrm{th}}$ Landau level \cite{Dots00}.
Both the effective potential obtained by this projection and the Fermi momenta
$k_{F\sigma}$ in the Landau quantized channels of the emitter are functions of
the applied magnetic field.  Through Eq.~(\ref{deltaexp}) this leads to the
strong $B$ dependence of the edge exponents $\gamma_\sigma$ observed in
experiments with 3D emitter.

\begin{figure}
\includegraphics[width=0.45\textwidth]{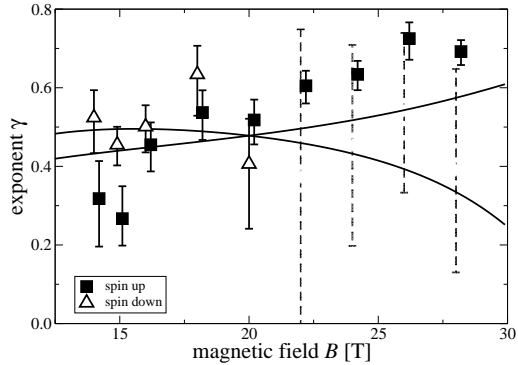}
\caption{
  \label{fig:exponents}
  Field dependence of the exponent $\gamma$ for both spin polarisations.  For
  $B>20\,T$ fitting of Eq.~(\ref{IVT}) to the FES for the minority spin is
  hampered by its decreasing weight.  This results in the large errors
  displayed for $\gamma_\downarrow$.  Full lines are theoretical predictions
  based on the phase shift (\ref{deltaexp}) of the exponentially screened
  scattering potential.  }
\end{figure}

In Ref.~\onlinecite{Dots00} the FES was analyzed based on the linear behaviour
of (\ref{deltaexp}) for small $k$.  In this approximation the difference of
the exponents $\gamma_\uparrow-\gamma_\downarrow$ is proportional to the
corresponding Fermi momenta which can be determined from the 1D DOS in the
Landau bands (broadened by disorder).  As a consequence, the difference of the
edge exponents should grow monotonically with $B$.  This behaviour is indeed
observed in the experiments at sufficiently high magnetic fields $B>B_c\approx
20\,T$.  Below $B_c$, however, the experimental data presented in
Ref.~\onlinecite{Dots00} indicate that $\gamma_\uparrow-\gamma_\downarrow$
changes sign, see also Fig.~\ref{fig:exponents}.
According to Eq.~(\ref{gamma}) such a crossing requires that
$\delta(k_{F\uparrow})=\delta(k_{F\downarrow})$ for $B_c$.  Clearly, such a
feature requires a non-monotonic $k$-dependence of the Fermi phase shift and
can only be captured when the full nonlinear expression (\ref{deltaexp}) for
the scattering phase is taken into account.  In Fig.~\ref{fig:exponents}, we
present a theoretical prediction based on this expression.  Both Landau level
broadening (affecting the field dependence of the Fermi momenta) and bending
of the Landau band in the emitter at $z=0$ have to be taken into account.  The
latter is found to be close to the threshold of formation of a single electron
bound state at the interface.  This is consistent with our picture that at
very high bias voltages a potential well is formed in the emitter close to the
barriers.  At the same time this limits the validity of the simple description
for the scattering potential in the emitter in terms of a decaying exponential
used in the derivation of Eq.~(\ref{deltaexp}).
Within this simple model the qualitative $B$-dependence of the experimental
data near $B_c$ is reproduced.  The splitting of the edge exponents at
$B<B_c$, however, is underestimated by the prediction.  Note that the
$z$-dependence of the scattering potential for the electrons in the emitter
can, in principle, be determined from the field dependence of the edge
exponents.  For this, however, they have to be extracted from the experimental
data with sufficient accuracy.  Here the quality of data is limited in
particular by the low weight of the FES for the minority spin $\downarrow$ at
fields $B\agt B_c$.

In this Paper we have analyzed FES observed in resonant tunneling experiments
through InAs quantum dots between 3D leads in strong magnetic fields parallel
to the current using both its temperature and bias-voltage dependence.  Based
on an explicit expression (\ref{IVT}) for the $I$-$V$ characteristic of the
FES data taken at different temperatures have been collapsed into a single
scaling form.  From these data the effect of the temperature-independent
fluctuations in the local density of states in the emitter on the current has
been identified.  Furthermore, the parameters describing the interaction
between the dot and the lead could be extracted.
Based on a microscopic model for the tunneling device the magnetic field
dependence of the broadening of the local level on the quantum dot has been
described.  In addition the exponent characterizing the FES for both spin
projections has been computed taking into account the non-linear momentum
dependence of the Fermi phase shift (\ref{deltaexp}), reproducing the
qualitative behaviour found in the experiment, in particular the change in
sign of $\gamma_\uparrow-\gamma_\downarrow$ at a magnetic field
$B\approx20\,T$.
For an improved description, both the effect of non-equilibrium
conditions\cite{MuAB03,AbLe05} has to be considered and fine-tuning the
scattering potential for the electrons in the emitter used to compute the
Fermi phase shift is necessary.  The latter would amount to introducing
introduction of additional fitting parameters, though.
On the other hand, we note that, this potential can be reconstructed in
principle from the experimental data provided that the edge exponents can be
extracted with sufficient accuracy.

\begin{acknowledgments}
This work has been supported by the Deutsche
Forschungsgemeinschaft and the BMBF.
\end{acknowledgments}


\begin{thebibliography}{25}
\expandafter\ifx\csname natexlab\endcsname\relax\def\natexlab#1{#1}\fi
\expandafter\ifx\csname bibnamefont\endcsname\relax
  \def\bibnamefont#1{#1}\fi
\expandafter\ifx\csname bibfnamefont\endcsname\relax
  \def\bibfnamefont#1{#1}\fi
\expandafter\ifx\csname citenamefont\endcsname\relax
  \def\citenamefont#1{#1}\fi
\expandafter\ifx\csname url\endcsname\relax
  \def\url#1{\texttt{#1}}\fi
\expandafter\ifx\csname urlprefix\endcsname\relax\def\urlprefix{URL }\fi
\providecommand{\bibinfo}[2]{#2}
\providecommand{\eprint}[2][]{\url{#2}}

\bibitem[{\citenamefont{Mahan}(1967)}]{Mahan67}
\bibinfo{author}{\bibfnamefont{G.~D.} \bibnamefont{Mahan}},
  \bibinfo{journal}{Phys. Rev.} \textbf{\bibinfo{volume}{163}},
  \bibinfo{pages}{612} (\bibinfo{year}{1967}).

\bibitem[{\citenamefont{Nozi{\`e}res and de~Dominicis}(1969)}]{NoDo69}
\bibinfo{author}{\bibfnamefont{P.}~\bibnamefont{Nozi{\`e}res}}
  \bibnamefont{and} \bibinfo{author}{\bibfnamefont{C.~T.}
  \bibnamefont{de~Dominicis}}, \bibinfo{journal}{Phys. Rev.}
  \textbf{\bibinfo{volume}{178}}, \bibinfo{pages}{1097} (\bibinfo{year}{1969}).

\bibitem[{\citenamefont{Matveev and Larkin}(1992)}]{MaLa92}
\bibinfo{author}{\bibfnamefont{K.~A.} \bibnamefont{Matveev}} \bibnamefont{and}
  \bibinfo{author}{\bibfnamefont{A.~I.} \bibnamefont{Larkin}},
  \bibinfo{journal}{Phys. Rev. B} \textbf{\bibinfo{volume}{46}},
  \bibinfo{pages}{15337} (\bibinfo{year}{1992}).

\bibitem[{\citenamefont{Geim et~al.}(1994)\citenamefont{Geim, Main, La~Scala,
  Eaves, Foster, Beton, Sakai, Sheard, Henni, Hill et~al.}}]{Geim94}
\bibinfo{author}{\bibfnamefont{A.~K.} \bibnamefont{Geim}},
  \bibinfo{author}{\bibfnamefont{P.~C.} \bibnamefont{Main}},
  \bibinfo{author}{\bibfnamefont{N.}~\bibnamefont{La~Scala},
  \bibfnamefont{Jr.}}, \bibinfo{author}{\bibfnamefont{L.}~\bibnamefont{Eaves}},
  \bibinfo{author}{\bibfnamefont{T.~J.} \bibnamefont{Foster}},
  \bibinfo{author}{\bibfnamefont{P.~H.} \bibnamefont{Beton}},
  \bibinfo{author}{\bibfnamefont{J.~W.} \bibnamefont{Sakai}},
  \bibinfo{author}{\bibfnamefont{F.~W.} \bibnamefont{Sheard}},
  \bibinfo{author}{\bibfnamefont{M.}~\bibnamefont{Henni}},
  \bibinfo{author}{\bibfnamefont{G.}~\bibnamefont{Hill}},  \bibnamefont{and}
  \bibinfo{author}{\bibfnamefont{M.~A.} \bibnamefont{Pate}},
  \bibinfo{journal}{Phys. Rev. Lett.} \textbf{\bibinfo{volume}{72}},
  \bibinfo{pages}{2061} (\bibinfo{year}{1994}).

\bibitem[{\citenamefont{Thornton
  et~al.}(1998{\natexlab{a}})\citenamefont{Thornton, Ihn, Main, Eaves,
  Benedict, and Henini}}]{Nott98a}
\bibinfo{author}{\bibfnamefont{A.~S.~G.} \bibnamefont{Thornton}},
  \bibinfo{author}{\bibfnamefont{T.}~\bibnamefont{Ihn}},
  \bibinfo{author}{\bibfnamefont{P.~C.} \bibnamefont{Main}},
  \bibinfo{author}{\bibfnamefont{L.}~\bibnamefont{Eaves}},
  \bibinfo{author}{\bibfnamefont{K.~A.} \bibnamefont{Benedict}},
  \bibnamefont{and} \bibinfo{author}{\bibfnamefont{M.}~\bibnamefont{Henini}},
  \bibinfo{journal}{Physica B} \textbf{\bibinfo{volume}{249-251}},
  \bibinfo{pages}{689} (\bibinfo{year}{1998}{\natexlab{a}}).

\bibitem[{\citenamefont{Benedict et~al.}(1998)\citenamefont{Benedict, Thornton,
  Ihn, Main, Eaves, and Henini}}]{Nott98b}
\bibinfo{author}{\bibfnamefont{K.~A.} \bibnamefont{Benedict}},
  \bibinfo{author}{\bibfnamefont{A.~S.~G.} \bibnamefont{Thornton}},
  \bibinfo{author}{\bibfnamefont{T.}~\bibnamefont{Ihn}},
  \bibinfo{author}{\bibfnamefont{P.~C.} \bibnamefont{Main}},
  \bibinfo{author}{\bibfnamefont{L.}~\bibnamefont{Eaves}}, \bibnamefont{and}
  \bibinfo{author}{\bibfnamefont{M.}~\bibnamefont{Henini}},
  \bibinfo{journal}{Physica B} \textbf{\bibinfo{volume}{256-258}},
  \bibinfo{pages}{519} (\bibinfo{year}{1998}).

\bibitem[{\citenamefont{Hapke-Wurst et~al.}(2000)\citenamefont{Hapke-Wurst,
  Zeitler, Frahm, Jansen, Haug, and Pierz}}]{Dots00}
\bibinfo{author}{\bibfnamefont{I.}~\bibnamefont{Hapke-Wurst}},
  \bibinfo{author}{\bibfnamefont{U.}~\bibnamefont{Zeitler}},
  \bibinfo{author}{\bibfnamefont{H.}~\bibnamefont{Frahm}},
  \bibinfo{author}{\bibfnamefont{A.~G.~M.} \bibnamefont{Jansen}},
  \bibinfo{author}{\bibfnamefont{R.~J.} \bibnamefont{Haug}}, \bibnamefont{and}
  \bibinfo{author}{\bibfnamefont{K.}~\bibnamefont{Pierz}},
  \bibinfo{journal}{Phys. Rev. B} \textbf{\bibinfo{volume}{62}},
  \bibinfo{pages}{12621} (\bibinfo{year}{2000}), \eprint{cond-mat/0003400}.

\bibitem[{\citenamefont{Gryglas et~al.}(2004)\citenamefont{Gryglas, Baj,
  Chenaud, Jouault, Cavanna, and Faini}}]{GryX04}
\bibinfo{author}{\bibfnamefont{M.}~\bibnamefont{Gryglas}},
  \bibinfo{author}{\bibfnamefont{M.}~\bibnamefont{Baj}},
  \bibinfo{author}{\bibfnamefont{B.}~\bibnamefont{Chenaud}},
  \bibinfo{author}{\bibfnamefont{B.}~\bibnamefont{Jouault}},
  \bibinfo{author}{\bibfnamefont{A.}~\bibnamefont{Cavanna}}, \bibnamefont{and}
  \bibinfo{author}{\bibfnamefont{G.}~\bibnamefont{Faini}},
  \bibinfo{journal}{Phys. Rev. B} \textbf{\bibinfo{volume}{69}},
  \bibinfo{pages}{165302} (\bibinfo{year}{2004}), \eprint{cond-mat/0402660}.

\bibitem[{\citenamefont{Khanin and Vdovin}(2005)}]{KhVd05}
\bibinfo{author}{\bibfnamefont{Y.~N.} \bibnamefont{Khanin}} \bibnamefont{and}
  \bibinfo{author}{\bibfnamefont{E.~E.} \bibnamefont{Vdovin}},
  \bibinfo{journal}{JETP Lett.} \textbf{\bibinfo{volume}{81}},
  \bibinfo{pages}{267} (\bibinfo{year}{2005}).

\bibitem[{\citenamefont{Schotte and Schotte}(1969)}]{Scho69}
\bibinfo{author}{\bibfnamefont{K.~D.} \bibnamefont{Schotte}} \bibnamefont{and}
  \bibinfo{author}{\bibfnamefont{U.}~\bibnamefont{Schotte}},
  \bibinfo{journal}{Phys. Rev.} \textbf{\bibinfo{volume}{182}},
  \bibinfo{pages}{479} (\bibinfo{year}{1969}).

\bibitem[{\citenamefont{Affleck and Ludwig}(1994)}]{AfLu94}
\bibinfo{author}{\bibfnamefont{I.}~\bibnamefont{Affleck}} \bibnamefont{and}
  \bibinfo{author}{\bibfnamefont{A.~W.~W.} \bibnamefont{Ludwig}},
  \bibinfo{journal}{J. Phys. A} \textbf{\bibinfo{volume}{27}},
  \bibinfo{pages}{5375} (\bibinfo{year}{1994}), \eprint{cond-mat/9405057}.

\bibitem[{\citenamefont{Bascones et~al.}(2000)\citenamefont{Bascones, Herrero,
  Guinea, and Sch{\"o}n}}]{BHGS00}
\bibinfo{author}{\bibfnamefont{E.}~\bibnamefont{Bascones}},
  \bibinfo{author}{\bibfnamefont{C.~P.} \bibnamefont{Herrero}},
  \bibinfo{author}{\bibfnamefont{F.}~\bibnamefont{Guinea}}, \bibnamefont{and}
  \bibinfo{author}{\bibfnamefont{G.}~\bibnamefont{Sch{\"o}n}},
  \bibinfo{journal}{Phys. Rev. B} \textbf{\bibinfo{volume}{61}},
  \bibinfo{pages}{16778} (\bibinfo{year}{2000}), \eprint{cond-mat/0002135}.

\bibitem[{\citenamefont{Muzykantskii et~al.}(2003)\citenamefont{Muzykantskii,
  d'Ambrumenil, and Braunecker}}]{MuAB03}
\bibinfo{author}{\bibfnamefont{B.}~\bibnamefont{Muzykantskii}},
  \bibinfo{author}{\bibfnamefont{N.}~\bibnamefont{d'Ambrumenil}},
  \bibnamefont{and}
  \bibinfo{author}{\bibfnamefont{B.}~\bibnamefont{Braunecker}},
  \bibinfo{journal}{Phys. Rev. Lett.} \textbf{\bibinfo{volume}{91}},
  \bibinfo{pages}{266602} (\bibinfo{year}{2003}), \eprint{cond-mat/0304583}.

\bibitem[{\citenamefont{Abanin and Levitov}(2005)}]{AbLe05}
\bibinfo{author}{\bibfnamefont{D.~A.} \bibnamefont{Abanin}} \bibnamefont{and}
  \bibinfo{author}{\bibfnamefont{L.~S.} \bibnamefont{Levitov}},
  \bibinfo{journal}{Phys. Rev. Lett.} \textbf{\bibinfo{volume}{94}},
  \bibinfo{pages}{186803} (\bibinfo{year}{2005}).

\bibitem[{\citenamefont{Ohtaka and Tanabe}(1984)}]{OhTa84}
\bibinfo{author}{\bibfnamefont{K.}~\bibnamefont{Ohtaka}} \bibnamefont{and}
  \bibinfo{author}{\bibfnamefont{Y.}~\bibnamefont{Tanabe}},
  \bibinfo{journal}{Phys. Rev. B} \textbf{\bibinfo{volume}{30}},
  \bibinfo{pages}{4235} (\bibinfo{year}{1984}).

\bibitem[{\citenamefont{Ohtaka and Tanabe}(1990)}]{OhTa90}
\bibinfo{author}{\bibfnamefont{K.}~\bibnamefont{Ohtaka}} \bibnamefont{and}
  \bibinfo{author}{\bibfnamefont{Y.}~\bibnamefont{Tanabe}},
  \bibinfo{journal}{Rev. Mod. Phys.} \textbf{\bibinfo{volume}{62}},
  \bibinfo{pages}{929} (\bibinfo{year}{1990}).

\bibitem[{\citenamefont{Yuval and Anderson}(1970)}]{YuAn70}
\bibinfo{author}{\bibfnamefont{G.}~\bibnamefont{Yuval}} \bibnamefont{and}
  \bibinfo{author}{\bibfnamefont{P.~W.} \bibnamefont{Anderson}},
  \bibinfo{journal}{Phys. Rev. B} \textbf{\bibinfo{volume}{1}},
  \bibinfo{pages}{1522} (\bibinfo{year}{1970}).

\bibitem[{\citenamefont{Gogolin et~al.}(1998)\citenamefont{Gogolin, Nersesyan,
  and Tsvelik}}]{GoNeTs:boson}
\bibinfo{author}{\bibfnamefont{A.~O.} \bibnamefont{Gogolin}},
  \bibinfo{author}{\bibfnamefont{A.~A.} \bibnamefont{Nersesyan}},
  \bibnamefont{and} \bibinfo{author}{\bibfnamefont{A.}~\bibnamefont{Tsvelik}},
  \emph{\bibinfo{title}{Bosonization and Strongly Correlated Systems}}
  (\bibinfo{publisher}{Cambridge University Press},
  \bibinfo{address}{Cambridge}, \bibinfo{year}{1998}).

\bibitem[{\citenamefont{Eggert et~al.}(1997)\citenamefont{Eggert, Mattsson, and
  Kinaret}}]{EgMK97}
\bibinfo{author}{\bibfnamefont{S.}~\bibnamefont{Eggert}},
  \bibinfo{author}{\bibfnamefont{A.~E.} \bibnamefont{Mattsson}},
  \bibnamefont{and} \bibinfo{author}{\bibfnamefont{J.~M.}
  \bibnamefont{Kinaret}}, \bibinfo{journal}{Phys. Rev. B}
  \textbf{\bibinfo{volume}{56}}, \bibinfo{pages}{R15537}
  (\bibinfo{year}{1997}), \eprint{cond-mat/9706157}.

\bibitem[{\citenamefont{Mattsson et~al.}(1997)\citenamefont{Mattsson, Eggert,
  and Johannesson}}]{MaEJ97}
\bibinfo{author}{\bibfnamefont{A.~E.} \bibnamefont{Mattsson}},
  \bibinfo{author}{\bibfnamefont{S.}~\bibnamefont{Eggert}}, \bibnamefont{and}
  \bibinfo{author}{\bibfnamefont{H.}~\bibnamefont{Johannesson}},
  \bibinfo{journal}{Phys. Rev. B} \textbf{\bibinfo{volume}{56}},
  \bibinfo{pages}{15615} (\bibinfo{year}{1997}), \eprint{cond-mat/9711204}.

\bibitem[{\citenamefont{Itskevich et~al.}(1996)\citenamefont{Itskevich, Ihn,
  Thornton, Henini, Foster, Moriarty, Nogaret, Beton, Eaves, and
  Main}}]{Nott96}
\bibinfo{author}{\bibfnamefont{I.~E.} \bibnamefont{Itskevich}},
  \bibinfo{author}{\bibfnamefont{T.}~\bibnamefont{Ihn}},
  \bibinfo{author}{\bibfnamefont{A.}~\bibnamefont{Thornton}},
  \bibinfo{author}{\bibfnamefont{M.}~\bibnamefont{Henini}},
  \bibinfo{author}{\bibfnamefont{T.~J.} \bibnamefont{Foster}},
  \bibinfo{author}{\bibfnamefont{P.}~\bibnamefont{Moriarty}},
  \bibinfo{author}{\bibfnamefont{A.}~\bibnamefont{Nogaret}},
  \bibinfo{author}{\bibfnamefont{P.~H.} \bibnamefont{Beton}},
  \bibinfo{author}{\bibfnamefont{L.}~\bibnamefont{Eaves}}, \bibnamefont{and}
  \bibinfo{author}{\bibfnamefont{P.~C.} \bibnamefont{Main}},
  \bibinfo{journal}{Phys. Rev. B} \textbf{\bibinfo{volume}{54}},
  \bibinfo{pages}{16401} (\bibinfo{year}{1996}).

\bibitem[{\citenamefont{Nauen et~al.}(2004)\citenamefont{Nauen, Hohls, Maire,
  Pierz, and Haug}}]{Nauen04}
\bibinfo{author}{\bibfnamefont{A.}~\bibnamefont{Nauen}},
  \bibinfo{author}{\bibfnamefont{F.}~\bibnamefont{Hohls}},
  \bibinfo{author}{\bibfnamefont{N.}~\bibnamefont{Maire}},
  \bibinfo{author}{\bibfnamefont{K.}~\bibnamefont{Pierz}}, \bibnamefont{and}
  \bibinfo{author}{\bibfnamefont{R.~J.} \bibnamefont{Haug}},
  \bibinfo{journal}{Phys. Rev. B} \textbf{\bibinfo{volume}{70}},
  \bibinfo{pages}{033305} (\bibinfo{year}{2004}).

\bibitem[{\citenamefont{Thornton
  et~al.}(1998{\natexlab{b}})\citenamefont{Thornton, Ihn, Main, Eaves, and
  Henini}}]{Nott98}
\bibinfo{author}{\bibfnamefont{A.~S.~G.} \bibnamefont{Thornton}},
  \bibinfo{author}{\bibfnamefont{T.}~\bibnamefont{Ihn}},
  \bibinfo{author}{\bibfnamefont{P.~C.} \bibnamefont{Main}},
  \bibinfo{author}{\bibfnamefont{L.}~\bibnamefont{Eaves}}, \bibnamefont{and}
  \bibinfo{author}{\bibfnamefont{M.}~\bibnamefont{Henini}},
  \bibinfo{journal}{Appl. Phys. Lett.} \textbf{\bibinfo{volume}{73}},
  \bibinfo{pages}{354} (\bibinfo{year}{1998}{\natexlab{b}}).

\bibitem[{\citenamefont{Westfahl et~al.}(1998)\citenamefont{Westfahl, Caldeira,
  Baeriswyl, and Miranda}}]{WCBM98}
\bibinfo{author}{\bibfnamefont{H.}~\bibnamefont{Westfahl}, \bibfnamefont{Jr.}},
  \bibinfo{author}{\bibfnamefont{A.~O.} \bibnamefont{Caldeira}},
  \bibinfo{author}{\bibfnamefont{D.}~\bibnamefont{Baeriswyl}},
  \bibnamefont{and} \bibinfo{author}{\bibfnamefont{E.}~\bibnamefont{Miranda}},
  \bibinfo{journal}{Phys. Rev. Lett.} \textbf{\bibinfo{volume}{80}},
  \bibinfo{pages}{2953} (\bibinfo{year}{1998}), \eprint{cond-mat/9710284}.

\bibitem[{\citenamefont{Main et~al.}(2000)\citenamefont{Main, Thornton, Hill,
  Stoddart, Ihn, Eaves, Benedict, and Henini}}]{Nott00}
\bibinfo{author}{\bibfnamefont{P.~C.} \bibnamefont{Main}},
  \bibinfo{author}{\bibfnamefont{A.~S.~G.} \bibnamefont{Thornton}},
  \bibinfo{author}{\bibfnamefont{R.~J.~A.} \bibnamefont{Hill}},
  \bibinfo{author}{\bibfnamefont{S.~T.} \bibnamefont{Stoddart}},
  \bibinfo{author}{\bibfnamefont{T.}~\bibnamefont{Ihn}},
  \bibinfo{author}{\bibfnamefont{L.}~\bibnamefont{Eaves}},
  \bibinfo{author}{\bibfnamefont{K.~A.} \bibnamefont{Benedict}},
  \bibnamefont{and} \bibinfo{author}{\bibfnamefont{M.}~\bibnamefont{Henini}},
  \bibinfo{journal}{Phys. Rev. Lett.} \textbf{\bibinfo{volume}{84}},
  \bibinfo{pages}{729} (\bibinfo{year}{2000}).

\end{thebibliography}

\end{document}